\documentclass[aps,twocolumn,prb,groupedaddress,showpacs]{revtex4}

\usepackage[utf8]{inputenc}
\usepackage{graphicx}
\graphicspath{{Figures_Nils/}}

\newcommand{\refq}[1]{(\ref{eq:#1})}
\newcommand{\nn}{n_{\rm{n}}}
\newcommand{\nw}{n_{\rm{w}}}
\newcommand{\reff}[1]{Fig.\ \ref{fig:#1}}
\newcommand{\reffa}[1]{Fig.\ \ref{fig:#1}(a)}
\newcommand{\reffb}[1]{Fig.\ \ref{fig:#1}(b)}
\newcommand{\reffc}[1]{Fig.\ \ref{fig:#1}(c)}
\newcommand{\reffd}[1]{Fig.\ \ref{fig:#1}(d)}

\newcommand{\refs}[1]{Sec.~\ref{sec:#1}}

\begin{document}

\title{Orbital-selective Mott transitions in a doped two-band Hubbard model\\ with crystal field splitting}

\author{Eberhard Jakobi}
\author{Nils Bl\"umer}
\email{Nils.Bluemer@uni-mainz.de}
\author{Peter van Dongen}
\affiliation{Institute of Physics, Johannes Gutenberg University, 55099 Mainz, Germany}
\date{\today}

\begin{abstract} 
We investigate the effects of crystal field splitting in a doped two-band Hubbard model with different bandwidths within dynamical mean-field theory (DMFT), using a quantum Monte Carlo impurity solver. 
In addition to an orbital-selective Mott phase (OSMP) of the {\it narrow} band, which is adiabatically connected with the well-studied OSMP in the half-filled case without crystal field splitting, we find, for sufficiently strong interaction and a suitable crystal field, also an OSMP of the {\it wide} band. We establish the phase diagram (in the absence of magnetic or orbital order) at moderate doping as a function of interaction strength and crystal field splitting and show that also the wide-band OSMP is associated with non-Fermi-liquid behavior in the case of Ising type Hund rule couplings.
Our numerical results are supplemented by analytical strong-coupling studies of spin order and spectral functions at integer filling.
\end{abstract}

\maketitle

\section{Introduction}\label{sec:intro}
As a fundamentally nonperturbative phenomenon beyond simple electronic band pictures, the Mott-Hubbard metal-insulator transition has been a subject of great interest in solid-state physics for decades.\cite{Gebhard97} In recent years, the Mott-Hubbard transition drew significant additional attention in the context of ultracold atoms.\cite{Bloch08_ferm,Esslinger08_Nature}
In the traditional scenario, merging the suggestions by Hubbard\cite{hubbard_hIII64} and Brinkmann and Rice,\cite{PhysRevB.2.4302} the electrons in a half-filled valence band acquire more and more effective mass with increasing interactions until they localize simultaneously and form a paramagnetic insulating state; this picture has found support in numerous calculations within dynamical mean-field theory\cite{PhysRevLett.62.324,georges:dmft96} (DMFT) for the Hubbard model\cite{gutzwiller:HubbardModel63,kanamori:HubbardModel,hubbard_hIII64} and its extensions to multiple\cite{PhysRevB.55.R4855} equivalent (e.g.,\ to three t$_{\text{2g}}$) orbitals. However, more recent experiments on ruthenates\cite{PhysRevLett.84.2666,PhysRevB.62.6458} have indicated that interesting complications can arise in the presence of multiple {\it inequivalent} valence orbitals with different effective bandwidths: then, as first illustrated in a simplistic model without inter-orbital coupling,\cite{EPJB.25.191}
increasing interactions $U>U_{c1}$ first localize the electrons in the narrow orbital(s) while the wide-band electrons initially remain itinerant and become insulating only at $U>U_{c2}>U_{c1}$. Subsequently, such orbital-selective Mott transitions (OSMTs) have been discussed for various classes of materials;\cite{Vojta_OSMT} the idea of partial localization has also been extended to ``momentum-selective Mott transitions'' in the context of high-$T_{\text{c}}$ materials.\cite{Gull2009,PhysRevB_80_064501}

While it was immediately clear that a realistic description would require at least three bands for the ruthenates\cite{EPJB.25.191} and that the mechanisms leading to different band widths would generically also affect the band center (i.e., induce a crystal field), most studies\cite{JPhysCondMat.19.436206,inaba:155106,de'medici:205124,PhysRevB.72.205126,PhysRevLett_99_236404,Arita2005} so far have addressed a minimal two-band Hubbard model
\begin{eqnarray}
 H_{\parallel} &=& -\sum_{\langle {\bf ij}\rangle m\sigma} t_m  c_{{\bf i}m\sigma}^\dagger  c_{{\bf j}m\sigma}^{\phantom{\dagger}}
+\; U\sum_{{\bf i}m} n_{{\bf i}m\uparrow}^{\phantom{\dagger}} n_{{\bf i}m\downarrow}^{\phantom{\dagger}} 
\nonumber \\
&&+ \sum_{{\bf i}\sigma\sigma'}\left(U' -\delta_{\sigma\sigma'}^{\phantom{\dagger}}J_z^{\phantom{\dagger}}\right) n_{{\bf i}1\sigma}^{\phantom{\dagger}} n_{{\bf i}2\sigma'}^{\phantom{\dagger}} \;,\label{eq:Jz}
\end{eqnarray}
in which the two orbitals $m\in\{1,2\}$ differ only by their hopping amplitude $t_m$ (between nearest-neighbor sites ${\bf i},{\bf j}$) and share the same local intra-orbital interaction $U$ and site potential; also the third term, containing both an inter-orbital Coulomb repulsion, parameterized by $U'$ (with $0< U' < U$), and an Ising-type Hund rule coupling with amplitude $J_z >0$, is symmetric in the orbital index. The DMFT studies of this model have almost exclusively assumed a semi-elliptic ``Bethe'' density of states and the absence of any magnetic order and mostly focused on half filling ($n=2$) and (to a lesser degree) on a hopping ratio of $t_1/t_2=2$. The expectation that this ``standard model'' for OSMTs captures the essence of the phenomenon, i.e., resolves {\em two} distinct orbital-selective transitions at half filling and low temperatures, could indeed be verified,\cite{knecht2005} after some initial confusion.\cite{liebsch:165103} In accordance with previous literature,\cite{PhysicaBCondMat.359.1366,pvd2006} we will refer to this model as the ``$J_z$-model''.  The general doped case was investigated in detail in Ref.\ \onlinecite{PhysRevB_80_115109} with the use of quantum Monte Carlo (QMC) methods and previously also by exact diagonalization\cite{koga:216402} and slave boson methods.\cite{Rueegg05}

In the following, we will explore the more general Hamiltonian
\begin{eqnarray}
 H_{\Delta} &=& H_{\parallel} + {\textstyle{\frac{1}{2}}}\Delta \sum_{{\bf i}\sigma} \left( n_{{\bf i}1\sigma}^{\phantom{\dagger}}-  n_{{\bf i}2\sigma}^{\phantom{\dagger}}\right) \label{eq:fullHamiltonian} \\
 && + {\textstyle{\frac{1}{2}}} J_{\perp}^{\phantom{\dagger}}\sum_{{\bf i}m\sigma}  c_{{\bf i}m\sigma}^\dagger \left(  c_{{\bf i}\bar{m}\bar{\sigma}}^\dagger  c_{{\bf i}m\bar{\sigma}}^{\phantom{\dagger}} + c_{{\bf i}m\bar{\sigma}}^\dagger  c_{{\bf i}\bar{m}\bar{\sigma}}^{\phantom{\dagger}} \right)  c_{{\bf i}\bar{m}\sigma}^{\phantom{\dagger}}\,.\nonumber
\end{eqnarray}
Here, the third term, proportional to $J_{\perp}$, describes spin flips and pair hopping processes arising from the general Hund rule coupling. The second term, proportional to $\Delta$, shifts the relative positions of the atomic energy levels of the two orbitals and, hence, describes crystal field splitting. First results regarding the impact of the latter term in the three-band extension of $H_{\Delta}$ have been obtained within a slave boson approximation.\cite{0611075}
Other studies have included crystal field terms as in \refq{fullHamiltonian}, but for orbitals with identical band widths.\cite{werner:126405,deMedici09,Huang2012}
Our goal in this paper is to explore the physics of $H_{\Delta}$ (with $t_1\not=t_2$ and $\Delta\not=0$) using QMC simulations within DMFT, identify new phases, and discuss spectral properties. An important special case of $H_{\Delta}$ occurs for $J_{\perp}=J_{z}\equiv J$, which we refer to as the ``$J$-model''. 

The $J$-model with $t_1=t_2$ and $\Delta =0$ is rotationally invariant in the sense of Castellani {\it et al.}\cite{PhysRevLett.43.1957}
We note, however, that the experimental systems allegedly described by $H_{\Delta}$, including the perovskite ruthenates, have less than cubic symmetry, so that there is no reason for assuming a rotationally symmetric screened Coulomb interaction in Eq.
 \refq{fullHamiltonian}, if the asymmetric hopping and the spin-orbit interaction are taken into account. Crystal field splitting of $d$ orbitals also reflects a broken rotational symmetry. For this physical reason, we will below set $J_{\perp}=0$ in QMC calculations. In analytical arguments we will assume $0\leq J_{\perp}\leq J_{z}$, which allows for antiferromagnetism at strong coupling. The choice $J_{\perp}=0$ in numerical simulations has the important additional advantage of avoiding a sign problem.

Note that, irrespective of the value of $J_{\perp}$, our Hamiltonian $H_{\Delta}$ is particle-hole symmetric only for $\Delta =0$; for $\Delta\neq 0$, it is mapped to $H_{-\Delta}$ under a particle-hole transformation. Accordingly, we can cover the entire density- and $\Delta$-range by assuming $n\geq 2$ and $\Delta\in \rm{I\hspace*{-.4ex}R}$; results for $n<2$ then follow from particle-hole symmetry. 

This paper is built up as follows. First, in \refs{analytics}, we consider the analytical properties of the $J_z$- and the $J$-models. In particular, we calculate the strong-coupling Hamiltonian and the local spectral functions at half filling; we also comment on the strong-coupling Hamiltonian at quarter filling. Then, in \refs{QMC}, we discuss the results of our QMC simulations for the $J_z$-model, in particular for orbital occupations, spectral functions and for the phase diagram. We also comment on non-Fermi-liquid properties on the basis of imaginary-frequency self-energy data. Generally, for the QMC calculations, we will concentrate on the question of metallicity as an effect of correlations; hence, we restrict ourselves to paramagnetic phases. We close (in \refs{summary}) with a summary and an outlook. Technical details regarding the strong-coupling expansion of the spectral functions are deferred to an appendix.


\section{Analytical results for the $J_z$- and $J$-models}\label{sec:analytics}

In this section we present analytical results for the $J_z$- and $J$-models with general $\Delta\neq 0$ for integer fillings at {\it strong\/} coupling ($U\rightarrow\infty$). Results of interest include the effective strong-coupling Hamiltonians at half- and quarter filling, which provide information about the symmetries and low-temperature phases of the models, and the spectral functions. We note that the model Hamiltonian $H_{\Delta}$ is $SU(2)$-symmetric in the spin sector for $J_{\perp}= J_{z}$; at $\Delta =0$ and $t_{1}=t_{2}$, it is also rotationally symmetric (i.e., $SO(2)$-symmetric) in the band index. In addition, the model with $\Delta =0$ is particle-hole symmetric at half filling. Crystal field splitting breaks the rotational symmetry for $J_{\perp}= J_{z}$ and the particle-hole symmetry at half filling.

\subsection{Strong-coupling Hamiltonian at half filling}

For the {\it half-filled\/} model $H_{\Delta}$ with $J_{\perp}= J_{z}$ and $\Delta =0$, an effective strong-coupling Hamiltonian was derived by Ferrero {\it et al.}\cite{PhysRevB.72.205126} Since most of their arguments are also valid for the model without particle-hole symmetry ($\Delta\neq 0$), and calculations for the model with $J_{\perp}< J_{z}$ are very similar, we only sketch the derivation. The effective strong-coupling Hamiltonian is obtained from standard Harris-Lange degenerate perturbation theory, which is based on a canonical transformation from the Hubbard electrons $c_{{\bf i}m\sigma}$ to new particles $\bar{c}_{{\bf i}m\sigma}$, whose hopping leaves the associated total number of double occupancies invariant (see also Appendix). For these new particles, we then define annihilation operators of double occupancies $d_{{\bf i}\sigma}=\bar{c}_{{\bf i}2\sigma}\bar{c}_{{\bf i}1\sigma}$ (with $\sigma =\uparrow ,\downarrow$) and $d_{{\bf i}0}=\frac{1}{\sqrt{2}}(\bar{c}_{{\bf i}2\downarrow}\bar{c}_{{\bf i}1\uparrow}+\bar{c}_{{\bf i}2\uparrow}\bar{c}_{{\bf i}1\downarrow})$, and a ($S=1$)-spin, constructed from these $d_{{\bf i}\sigma}$- and $d_{{\bf i}0}$-operators:
\[
{\bf S}_{\bf i}=\left(\begin{array}{c} \frac{1}{\sqrt{2}}(d^{\dagger}_{{\bf i}\uparrow}d^{\phantom{\dagger}}_{{\bf i}0} + d^{\dagger}_{{\bf i}0}d^{\phantom{\dagger}}_{{\bf i}\uparrow} + d^{\dagger}_{{\bf i}\downarrow}d^{\phantom{\dagger}}_{{\bf i}0} + d^{\dagger}_{{\bf i}0}d^{\phantom{\dagger}}_{{\bf i}\downarrow}) \\ \frac{i}{\sqrt{2}}(d^{\dagger}_{{\bf i}0}d^{\phantom{\dagger}}_{{\bf i}\uparrow} - d^{\dagger}_{{\bf i}\uparrow}d^{\phantom{\dagger}}_{{\bf i}0} + d^{\dagger}_{{\bf i}\downarrow}d^{\phantom{\dagger}}_{{\bf i}0} - d^{\dagger}_{{\bf i}0}d^{\phantom{\dagger}}_{{\bf i}\downarrow}) \\ d^{\dagger}_{{\bf i}\uparrow}d^{\phantom{\dagger}}_{{\bf i}\uparrow} - d^{\dagger}_{{\bf i}\downarrow}d^{\phantom{\dagger}}_{{\bf i}\downarrow}\end{array}\right)\; .
\]
The relevance of the $d_{{\bf i}\sigma}$- and $d_{{\bf i}0}$-operators is that the three atomic states $d^{\dagger}_{{\bf i}\uparrow}|0\rangle$, $d^{\dagger}_{{\bf i}\downarrow}|0\rangle$ and $d^{\dagger}_{{\bf i}0}|0\rangle$ are lowest in energy (and, hence, span a triplet) for $J_{\perp}=J_{z}\equiv J$, while the two states $d^{\dagger}_{{\bf i}\uparrow}|0\rangle$ and $d^{\dagger}_{{\bf i}\downarrow}|0\rangle$ are lowest in energy if $0\leq J_{\perp}<J_{z}$. With these definitions, the effective $S=1$ spin Hamiltonian for $J_{\perp}=J_{z}= J$ is given by:
\[
H_{t}' =\sum_{\langle {\bf ij}\rangle} J^{\phantom{\dagger}}_{\rm Heis}({\bf S}_{\bf i}\cdot {\bf S}_{\bf j} -n_{\bf i}\, n_{\bf j})\;\; ,\;\; J^{\phantom{\dagger}}_{\rm Heis}\equiv \frac{(t_{1})^{2}+(t_{2})^{2}}{U+J} 
\]
where the number operator is defined as $n_{\bf i}\equiv\sum_{\sigma}d^{\dagger}_{{\bf i}\sigma}d^{\phantom{\dagger}}_{{\bf i}\sigma}+d^{\dagger}_{{\bf i}0}d^{\phantom{\dagger}}_{{\bf i}0}$. The effective Hamiltonian for $0\leq J_{\perp}<J_{z}$ reads:
\[ 
H_{t}' =\sum_{\langle {\bf ij}\rangle} J^{\phantom{\dagger}}_{\rm Is}(S_{{\bf i}3}\, S_{{\bf j}3} -n_{\bf i}\, n_{\bf j})\;\; ,\;\; J^{\phantom{\dagger}}_{\rm Is}=\frac{(t_{1})^{2}+(t_{2})^{2}}{U+J_{z}} \; ,
\]
where now $n_{\bf i}\equiv\sum_{\sigma}d^{\dagger}_{{\bf i}\sigma}d^{\phantom{\dagger}}_{{\bf i}\sigma}$. Note that the value of $J_\perp$ becomes irrelevant if $J_z >J_\perp$. These results are generally valid in any spatial dimension. 

It is interesting to note that the results stated above for the effective Hamiltonians at strong coupling are entirely {\it independent\/} of the crystal field splitting parameter $\Delta$, although they are valid for general $\Delta <U+J_{z}$ if $J_{\perp}\leq J_{z}$. The explanation is, technically, that the sum of the excitation energies for hopping processes, starting from the subspace spanned by $d^{\dagger}_{{\bf i}\uparrow}|0\rangle$, $d^{\dagger}_{{\bf i}\downarrow}|0\rangle$ and possibly $d^{\dagger}_{{\bf i}0}|0\rangle$, is independent of $\Delta$. This, in turn, occurs since the crystal field splitting term in the Hamiltonian commutes with the hopping.

In the limit of high spatial dimensions, the spin Hamiltonians, thus derived, are solved by mean field theory. In particular, the ``antiferromagnetic'' critical temperatures for these models can easily be calculated for a bipartite lattice in the limit of infinite coordination number $Z$:
\[
\frac{k^{\phantom{\dagger}}_{\hspace*{.2ex}{\rm B}}T^{\phantom{\dagger}}_{\rm c}}{Z} \stackrel{Z\to \infty}{\longrightarrow}
\left\{
\begin{array}{ll} 
	J^{\phantom{\dagger}}_{\rm Heis}&\text{ for } J_{\perp}=J_{z}\equiv J\\
	J^{\phantom{\dagger}}_{\rm Is} & \text{ for } J_{\perp}<J_{z}\,.
\end{array}	
\right.
\]
These results are interesting, because they show that $T^{\phantom{\dagger}}_{\rm c}$ is dominated in both models by the largest hopping amplitude, which is that of the broad band ($t_{2}^{*}\equiv t_2\sqrt{Z}$). We conclude that the {\it broad band\/} primarily determines the energy scale at strong coupling; from previous work\cite{VanDongen2007} we know that the energy scales of antiferromagnetism at weak coupling are primarily determined by the {\it narrow band\/} ($t_{1}^{*}\equiv t_1\sqrt{Z}$). Note that, since the effective strong-coupling Hamiltonians are $\Delta$-independent (within their range of validity), the same holds for the critical temperatures.

\subsection{The spectral functions at half filling}

The calculation of the spectral functions on the Bethe lattice (which is of interest here) at strong coupling, $T=0$, and half filling proceeds along the lines of Kalinowski and Gebhard.\cite{Kalinowski2002} These authors calculated the spectral function for a {\it single\/}-band Hubbard model in strong-coupling perturbation theory. The analogous calculations for the {\it two-band\/} model $H_{\Delta}$, considered here, are technically considerably more complicated and are, hence, summarized in the Appendix. The results are, in terms of the noninteracting density of states $\nu^0_m(\omega)$:
\begin{eqnarray*}
\nu_{m\sigma}^{\rm LHB}(\omega )\!&=&\!\frac{1}{\sqrt{2}}\, \nu^0_m\Big(\!\sqrt{2}\, 
\Big[\omega +\frac{U\!+J_{z}\mp\Delta}{2}\Big]\Big) \quad (J_{\perp}<J_{z}) \\
\nu_{m\sigma}^{\rm LHB}(\omega )\!&=&\!\frac{1}{\sqrt{3}}\, \nu^0_m\Big(\!\sqrt{\frac{4}{3}}\, 
\Big[\omega +\frac{U\!+J\mp\Delta}{2}\Big]\Big) \quad (J_{\perp}=J_{z}) \\
\end{eqnarray*}
Here the upper sign refers to the $(m=1)$- and the lower sign to the $(m=2)$-orbital. It is interesting to note that the spectral function for $J_{\perp}=J_{z}\equiv J$ is {\it broader\/} than that for $J_{\perp}<J_{z}$ by a factor of $\sqrt{3/2}$, which reflects the larger number of possible hopping processes in the former case as compared to the latter. Clearly, if $J_{z}-J_{\perp}$ is positive but small, there will be a crossover from the spectrum for $J_{z}>J_{\perp}$ to that for $J_{z}=J_{\perp}$ at finite temperatures ($T>0$) or finite Hubbard interaction ($U<\infty$). The relevant temperature and $U^{-1}$-scales are $(J_{z}-J_{\perp})/k^{\phantom{\dagger}}_{\hspace*{.2ex}{\rm B}}$ and $(J_{z}-J_{\perp})/(t_{1,2}^{*})^{2}$, respectively.

We note that the effect of crystal field splitting at strong coupling is simply to shift the $(m=1)$-band energetically to the right and the $(m=2)$-band to the left. We will see below from the results of the QMC-simulations, that the effect of crystal field splitting at finite interaction strength is not quite so obvious.

\subsection{Strong-coupling Hamiltonian at quarter filling}

The determination of the strong-coupling Hamiltonian at quarter  or three-quarter filling is extremely simple, although the result is nontrivial. For example, in the atomic limit ($t_1 =t_2 =0$), the minimal energy at {\it quarter filling\/} is obtained for an ensemble of singly-occupied sites (if $|J_z|$ and $|J_\perp|$ are both smaller than $U$ and $U'$). 
The only relevant term left in the Hamiltonian is, therefore, the crystal field splitting ${\textstyle{\frac{1}{2}}}\Delta \sum_{{\bf i}\sigma} ( n_{{\bf i}1\sigma}^{\phantom{\dagger}}-n_{{\bf i}2\sigma}^{\phantom{\dagger}})$, implying that the ground state occurs for an ensemble of single occupancies in band 2 if $\Delta >0$ or, alternatively, single occupancies in band 1 if $\Delta <0$. The spins of these single occupancies are not fixed yet. Since all particles at quarter filling occupy one single band, we conclude that, if a small hopping of the particles is now switched on, the two-band Hamiltonian $H_{\Delta}$ at quarter filling reduces to a half-filled single-band Hubbard model. At strong coupling, the half-filled single-band Hubbard model reduces to an antiferromagnetic Heisenberg model. Hence we conclude that, at low temperatures, the quarter-filled two-band model $H_{\Delta}$ describes an orbitally ferromagnetic Heisenberg spin-antiferromagnet. 

We note that $H_{\Delta}$ at three-quarter filling is mapped by a particle-hole transformation to $H_{-\Delta}$ at quarter filling; the low-temperature physics is, again, that of an orbitally ferromagnetic Heisenberg spin-antiferromagnet.

These strong-coupling arguments assume that the bandwidths $W_1$ and $W_2$ of the two-band model are small compared to all other parameters, so that, in particular, $|\Delta | >W_{1,2}$. If this condition is not fulfilled, the results may change completely. For instance, in the special case $\Delta =0$ the two-band model at quarter filling changes its behavior both in the spin and in the orbital degrees of freedom and reduces to a spin-ferromagnetic Heisenberg orbital-antiferromagnet.\cite{Peters2010} The numerical results show that the system, in the ground state, may or may not have orbital long-range order, depending on the model parameters.


\section{QMC results for the $J_z$-model}\label{sec:QMC}
In this section, we present Hirsch-Fye QMC\cite{PhysRevLett.56.2521,Blumer2007} results for the $J_z$-model \refq{Jz} with $J_z = U/4$ and $U' = U/2$.
These values are consistent with estimates\cite{Noh2005} of $J=0.5 \text{ eV}$ and (multiplet-averaged) $U_{dd}=2.0 \text{ eV}$ for Ca$_{2-x}$Sr$_x$RuO$_4$;%
\footnote{Earlier studies had assumed\cite{Mizokawa2001} values of $J=0.5 \text{ eV}$ and $U_{dd}=3.0 \text{ eV}$ for Ca$_2$RuO$_4$; a similar fraction $J/U=0.15$ has also recently been used\cite{Yi2013} for modeling OSMT physics in A$_x$Fe$_{2-y}$Se$_2$ superconductors (where A=K, Rb).}
$J/U=1/4$ is also in the middle of the interval $0<J/U<1/2$ following from the relation $U'+2J=U$ (strictly valid only for cubic symmetry) under the natural assumption $0<U'<U$.
We will assume half-elliptic ``Bethe'' densities of states\cite{Kollar2005} for both bands, with a full bandwidth $W_\mathrm{n} = 2$ for the ``narrow'' band and 
$W_\mathrm{w} = 4$ for the ``wide'' band, respectively; this corresponds to the best studied case for $\Delta=0$. Also in line with earlier work, we will restrict ourselves  to the paramagnetic case, i.e., exclude antiferromagnetic and orbital order; the results should be relevant at intermediate temperatures or for frustrated systems. 
\footnote{Very recently, it has been shown that orbital-specific frustration can also lead to orbital-selective insulating phases.\cite{Lee2011a}}
In the following, we present QMC results obtained at a temperature $T = 1/40$ with a discretization parameter in the Trotter decomposition\cite{trotter_1959} of $\Delta\tau = 0.4$; these data were checked by additional calculations at different values of $\Delta\tau$.

In the following, we will first discuss orbital-specific occupation numbers and spectral functions, then construct the phase diagram and, finally, address possible non-Fermi-liquid properties on the basis of self-energy data, all as a function of crystal field splitting.

\subsection{Orbital occupation numbers}

\begin{figure}
\includegraphics[width=\columnwidth]{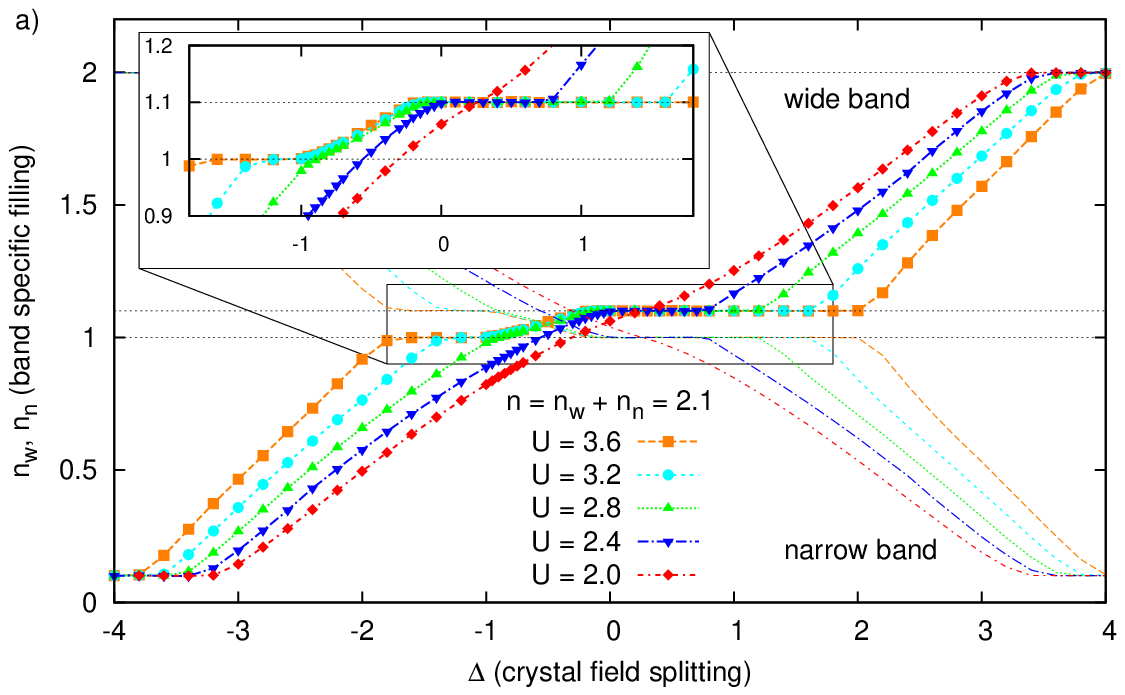}

\vspace{1ex}
\includegraphics[width=\columnwidth]{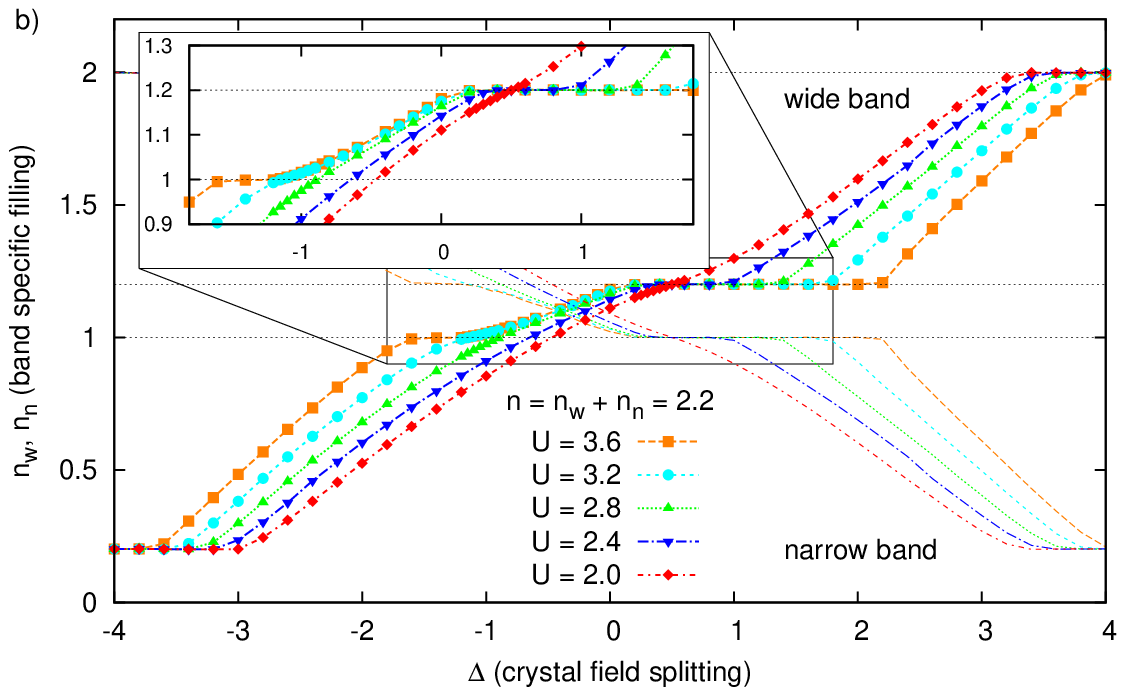}
\caption{(Color online) Orbital occupation numbers $\nn$, $\nw$ as a function of crystal field splitting $\Delta$: (a) for total filling  $n=2.1$ and (b) for $n=2.2$. Results for the wide (narrow) band are indicated by thick lines and symbols (thin lines). Mott plateaus in $\nn$, $\nw$ appear for $U>2$ (see magnified insets for $\nw$). \label{fig:filling}}
\end{figure}

As the crystal field splitting $\Delta$ acts like a magnetic field in the orbital sector, i.e., shifts all energy levels of the wide band downwards for $\Delta>0$ and all energy levels of the narrow band upwards, one expects that, generically, an increase of $\Delta$ will increase the filling in the wide band and decrease the filling in the narrow band. This applies both at constant chemical potential $\mu$ and at constant total filling $n=\nn+\nw$, unless one or both of the bands are incompressible, i.e., in a Mott or band insulating state. 

Only at integer filling, both bands can be incompressible at the same time: indeed, the model is known to be fully insulating at $n=2$, $U\gtrsim 2.8$, and $\Delta=0$;\cite{knecht2005} due to the first-order character of the associated transition, the insulating state must be stable also at small $\Delta$ (up to the order of the gap in the particle-hole symmetric case). At very large crystal field splitting ($|\Delta|\gtrsim \mathrm{max} \{W_{\rm w},W_{\rm n},U,k_B\,T\}$), one expects that both orbitals become band-insulating, with all electrons in the lower band (wide or narrow, depending on the sign of $\Delta$).

Away from half filling (here we concentrate on $n>2$), at least one of the bands must have a noninteger occupation and, therefore, remain metallic. This case appears much more interesting, so we will focus on it in the following. In \reffa{filling}, we present data for $n=2.1$, i.e., for relatively small electron doping. The case of larger doping will be discussed later [cf.\ \reffb{filling}] while results for hole doping ($n<2$) follow from symmetry. 
As expected, the occupation $\nw$ of the wide band (symbols and solid lines) increases monotonically with $\Delta$ at all interactions $2\le U \le 3.6$. Accordingly, the occupation $\nn=n-\nw$ decreases monotonically. 

Plateaus in the orbital occupations are observed for large absolute values of the crystal field splitting at all interactions beyond thresholds that increase with $U$, e.g., for $\Delta\gtrsim 4.0$ and for $\Delta\lesssim -3.8$ at $U=3.6$ (squares), which correspond to a totally filled wide and narrow band, respectively.  In between these (orbital selective) band insulating phases, the density curve $\nw(\Delta)$ is smooth with strictly positive slope at $U\lesssim 2.0$ (diamonds), indicating a purely metallic phase. 

At $U=2.4$ (triangles), an additional plateau appears at $0\lesssim \Delta\lesssim 0.8$.  
On this plateau, the occupation of the narrow band is integer ($n_{\rm n} = 1.0$) and that of the wide band fractional ($n_{\rm w}=1.1$). Since $\Delta$ acts as an orbital-dependent chemical potential (in addition to $\mu$), we can interpret the pinning of the narrow-band occupation at half filling as arising from an incompressibility of this subsystem; thus, the system appears as a narrow-band orbital-selective Mott insulator. In contrast, the ``pinning'' of $\nw$ at the value 1.1 arises from our constraint $n=\nn+\nw=2.1$; it would not show up in plots at constant $\mu$. The narrow-band Mott plateau broadens as the interaction is increased and extends to slightly negative $\Delta$ for $U> 2.4$. Thus, this phase can be identified with the one previously studied\cite{PhysRevB_80_115109} at $\Delta=0$ for $n=2.1$, which is continuously connected with the OSMP at $\Delta=0$ and half filling ($n=2$).\cite{knecht2005}

At stronger coupling $U>2.8$, additional plateaus appear at $\Delta\lesssim -1.0$. In this case, the {\it wide} band is half-filled and (by the above arguments) insulating. Evidently, the crystal field splitting must be essential for this wide-band OSMP since it cannot occur at $\Delta=0$ (when all phases with insulating bands are connected to half filling, i.e., particle-hole symmetry). It will, thus, be instructive to compare the two different types of OSMPs in some detail, in particular regarding spectral properties.

Before doing that, let us first discuss the impact of the doping level on the basis of \reffb{filling} which shows results for twice the doping strength (filling $n=2.2$) compared to \reffa{filling}. Overall, the results in both panels of \reff{filling} look very similar, up to slight shifts in critical values of $\Delta$ and, of course, in the noninteger plateau values. We conclude that the specific doping level is not crucial so that we can focus on the specific filling $n=2.1$ for the remainder of the paper without loss of generality.

\subsection{Spectral functions}

\begin{figure}
\includegraphics[width=\columnwidth]{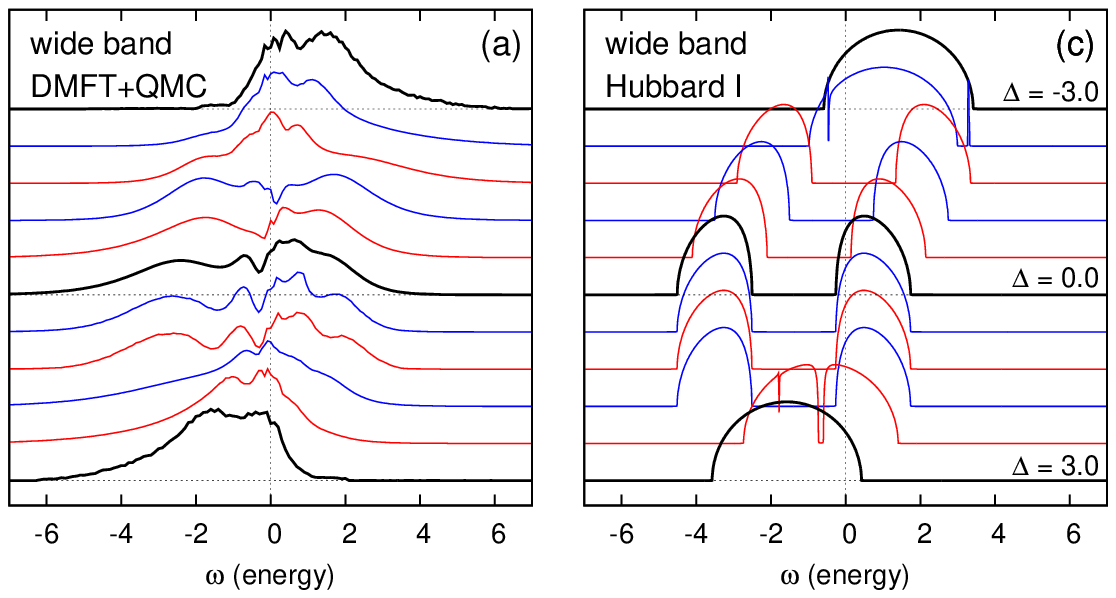}
\includegraphics[width=\columnwidth]{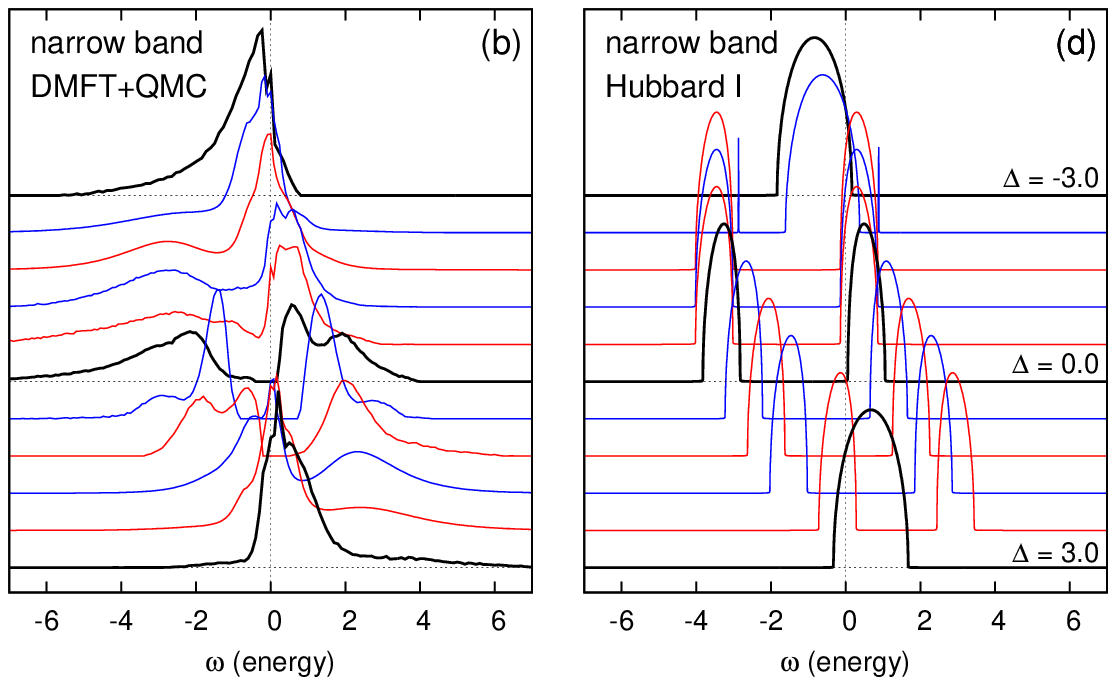}
\caption{(Color online) Spectral functions $A(\omega)$ for $n=2.1$ and $U=3.0$, with data for the wide (narrow) band shown in the upper (lower) panels. The QMC estimates (with maximum entropy analytic continuation), shown in the left column, indicate orbital-selective Mott phases of the narrow and (less clearly) of the wide band; corresponding Hubbard-I spectra capture mainly the peak positions (see text).
}\label{fig:spectra}
\end{figure}

Now we turn to the spectral properties, starting with numerical estimates, obtained by maximum entropy analytic continuation of QMC imaginary-time Green functions. Results for the wide band are shown in \reffa{spectra} and those for the narrow band in \reffb{spectra}; corresponding analytic results (right hand panels of \reff{spectra}) will be discussed later. As expected, the spectral weight of the wide band shifts towards smaller $\omega$ when the crystal field is increased from $\Delta=-3.0$ [topmost curve in \reffa{spectra}] to $\Delta=3.0$ [lowest curve in \reffa{spectra}], with a shape that changes significantly at intermediate values of $\Delta$ (shown with a spacing of 0.6). Conversely, the narrow-band spectral weight shifts upwards. 

The narrow-band insulating phase is clearly apparent as a gap around the Fermi energy for $0.0 \lesssim \Delta \lesssim 1.2$ in \reffb{spectra}, in line with the expectations from the orbital occupancy analysis of \reffa{filling}. This gap (with a maximum width of about 1.5) shifts with $\Delta$ roughly like the center of mass. More generally, the narrow-band spectrum starts out from a narrow shape, with a peak just below the Fermi energy at $\Delta=-3.0$, which moves towards larger $\omega$ with increasing $\Delta$ (and decays slowly for $\Delta\to 3.0$). At $\Delta\approx-1.8$, a second peak emerges at $\omega\approx -3$, the position of which is initially nearly frozen, then moves towards the gap edge at $\Delta\gtrsim 0$, until it becomes the main peak. At $\Delta=0.6$, the narrow-band spectrum is remarkably symmetric. Minor structures visible in the numerical results, such as a splitting of peaks for specific values of $\Delta$, are probably not significant.

Due to such numerical noise, the wide-band insulating phase is much harder to detect in the corresponding spectra, \reffa{spectra}. However, a well-developed dip is seen to cross the Fermi energy in the range $-1.2\le\Delta\le -0.6$, which is consistent with the very small gap at $\Delta\approx -1.0$ that one would expect from interpolating the occupancy data of \reffa{filling}. The dip persists at $\omega\approx -0.2$ for $\Delta\lesssim 1.2$, which might be interpreted as a sign of non-Fermi-liquid behavior (cf. Ref.\ \onlinecite{PhysRevB_80_115109}).

In \reffc{spectra} and \reffd{spectra} we show, for comparison, spectra obtained from a Hubbard-I-type solution of the same Hamiltonian. Obviously, this simple approximation, which does not include life time effects, cannot capture the highly nontrivial correlation physics and reproduce the corresponding spectra on a quantitative level. In particular, the Hubbard-I predictions contain unphysical or much too broad gaps. 
However, at strong crystal field splitting $|\Delta|=3.0$, the Hubbard-I line shapes are roughly correct.
In addition, the peak positions and their nonuniform evolution as a function of $\Delta$ agree with the QMC data remarkably well at all values of $\Delta$, which clearly supports the reliability of our numerical methods. For the narrow band, the insulating phase (gap around Fermi energy) is also predicted nearly correctly (with a false positive only for $\Delta=1.8$); in contrast, the wide-band insulating phase is grossly overestimated.

\subsection{Phase diagram}

\begin{figure} 
\includegraphics[width=\columnwidth]{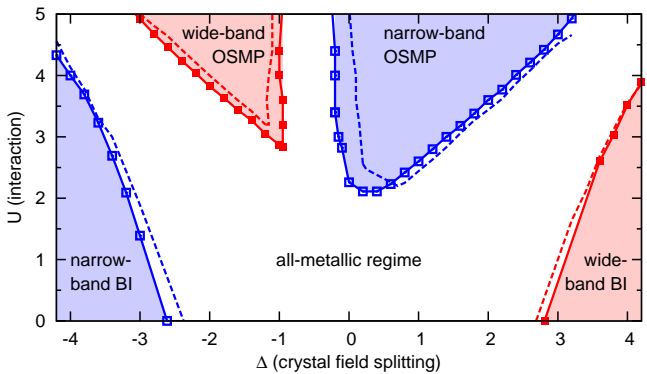}\\
\caption{(Color online) Phase diagram as a function of the interaction $U$ and the crystal field splitting $\Delta$ for $n=2.1$ (symbols, solid lines, and shaded regions) and $n=2.2$ (dashed lines). For large $|\Delta|$, the energetically lower of both orbitals is in a band insulating state (BI), while the other orbital contains the remaining electrons. Orbital-selective Mott phases (OSMPs) occur at sufficiently strong coupling ($U\gtrsim 2$) for moderate $|\Delta|$.}\label{fig:phase_diagram_cf}
\end{figure}

Using primarily orbital occupation data for a broad range of parameters $\Delta$ and $U$, we have constructed the phase diagram \reff{phase_diagram_cf}.
Specifically, the critical interactions for the onset of plateaus, i.e., phases where one of the orbitals is incompressible, were determined on a fine grid of $\Delta$ values. The resulting boundaries towards these orbital-selective insulating phases are shown as symbols and solid lines in \reff{phase_diagram_cf} for a total filling $n=2.1$ (while dashed lines denote transitions at $n=2.2$). At small $U\lesssim 2$, the system is either fully metallic or band insulating (in the narrow band for $\Delta \lesssim -3$, in the wide band for $\Delta \gtrsim 3$); the precise critical values of $\Delta$ shift significantly with $U$. At $U\gtrsim 2$, a narrow-band orbital-selective Mott phase emerges at $\Delta\gtrsim 0$. Only at $U\gtrsim 2.8$, an additional wide-band OSMP becomes stable at $\Delta \lesssim -1$. It is clear that the two OSMPs must be separated (at constant filling $n>2$) by a fully metallic region, where both orbitals are slightly more than half filled [cf. \reffa{filling}]; the extent of this ``finger'' of the all-metallic phase at $-1.0\lesssim \Delta \lesssim -0.2$ (at $n=2.1$) is practically independent of $U$, at least in the interaction range covered in this study.

Stronger doping ($n=2.2$, dashed lines in \reff{phase_diagram_cf}) suppresses the narrow-band OSMP at $\Delta\approx 0$, i.e., shifts the phase boundary to larger values of $U$ and broadens the central all-metallic phase. Otherwise, the effects are surprisingly small, which justifies, again, our focus on the single doping level $n=2.1$.

\subsection{Non-Fermi-liquid properties}\label{NON-FL-CF}

In order to get more insight into the nature of the orbital-selective phases shown in \reff{phase_diagram_cf}, beyond the orbital-selective incompressibility apparent in \reff{filling} and the spectral information of \reff{spectra}, let us now discuss signatures in the self-energy $\Sigma(\omega)$, a direct quantitative measure of correlation effects. Specifically, we consider the self-energy on the imaginary axis [i.e., for $\omega=i \omega_n=i (2n+1) \pi T$], which is directly available as a state variable in the DMFT self-consistency cycle, and avoid an ill-conditioned analytic continuation to the real axis (which limits the reliability of the spectral functions $A(\omega)=-\text{Im} G(\omega+i0^{+})/\pi$ shown in \reff{spectra}). 

\begin{figure} 
\includegraphics[width=\columnwidth]{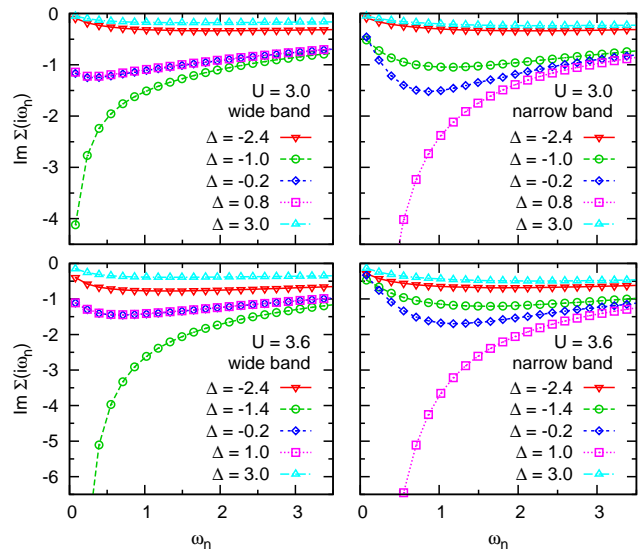}
\caption{(Color online) Imaginary part of self-energy for total filling $n=2.1$, interaction strengths $U=3.0$ (top), $U=3.6$ (bottom), and selected values of the crystal field splitting parameter $\Delta$.}\label{fig:SE}
\end{figure}

The upper panels of \reff{SE} show $\text{Im} \Sigma(i\omega_n)$ for the wide band (top left) and the narrow band (top right) at $U=3.0$. Deep in the fully metallic phase, at $\Delta=-2.4$ and at $\Delta=3.0$ (triangles), both observables have small absolute values and decay essentially to zero at small $\omega_n$, indicating a weakly correlated Fermi liquid. 

In the narrow-band insulating phase, at $\Delta=0.8$ (squares), the narrow-band self-energy diverges at small frequencies; the corresponding wide-band self-energy is nearly flat and tends to a finite value at small $\omega$. This clear sign of non-Fermi liquid behavior in the metallic component of an orbital-selective Mott phase appears completely analogous to the well-known situation at $n=2$ and $\Delta=0$.\cite{knecht2005} Quite remarkably, almost identical values of $\text{Im} \Sigma_{\text{w}}(i\omega_n)$ are obtained also at $\Delta=-0.2$, very close to the edge of the narrow-band OSMP, where the narrow-band self-energy $\text{Im} \Sigma_{\text{n}}(i\omega_n)$ is still strongly enhanced at intermediate $\omega_n\approx 1$, but decays almost Fermi liquid like at small $\omega_n$. 

In the wide-band insulating phase, at $\Delta=-1.0$ (circles), the role of the two bands is just exchanged, relative to the case discussed above: now the wide band shows a divergent self-energy, while the narrow band displays non-Fermi liquid behavior. We conclude that this behavior is really generic of itinerant electrons coupled via an Ising type Hund rule interaction to localized electrons and does not depend on details of the model.

A further indication that the behavior discussed so far is quite generic is the complete qualitative agreement between the self-energy data discussed so far for $U=3.0$ (top panels in \reff{SE}) and corresponding data at stronger interaction $U=3.6$ (bottom panels in \reff{SE}): up to a slight enhancement (note the change of scales between the two rows of panels), the imaginary parts of the self-energies on the Matsubara axis are nearly identical at corresponding phase points. The main difference is a much larger residual value at small frequencies in the fully metallic phases (triangles); apparently, at this stronger interaction (and close to the wide-band OSMP), the temperature $T=1/40$ is already above the range of ``good'' Fermi liquid behavior at $\Delta=-2.4$.

\section{Conclusion}\label{sec:summary}

In summary, we have investigated the effects of crystal field splitting $\Delta$ in the doped two-band model \refq{fullHamiltonian}
with a band width ratio of $t_{\text{w}}/t_{\text{n}}=2$ and Ising-type exchange interaction.
Using the Hirsch-Fye quantum Monte Carlo technique, we have calculated orbital filling factors, spectra, and Matsubara self-energies within DMFT, focusing on Mott physics within the spatially homogeneous phase.  
The resulting phase diagram contains not only an orbital-selective Mott phase of the narrow, i.e., generically more strongly correlated, band, but also a wide-band OSMP at suitable values of the crystal field (and strong enough interaction). This shows that the crystal field, i.e., a diagonal element in the tight-binding Hamiltonian, can be as relevant for Mott physics as the hopping, i.e., offdiagonal elements. 
Clear signatures of non-Fermi-liquid behavior are seen in the Matsubara self-energies in both types of OSMPs; these findings are also consistent with the complex evolution of the local spectral functions as a function of $\Delta$ that we observed using QMC and the maximum entropy method.

On the one hand, the QMC results presented in this paper complete the picture regarding the impact of a band width difference on the correlation physics of multiorbital systems. They show that the OSMT scenario established in earlier studies is not only stable with respect to doping,\cite{PhysRevB_80_115109} but also with respect to (additional) crystal field splitting. On the other hand, our finding of a wide-band OSMP
makes contact with earlier studies of the isolated impact of crystal field splitting on otherwise equivalent orbitals\cite{werner:126405,deMedici09,Huang2012} and shows that the combined effect of filling control and crystal field\cite{werner:126405} can drive an OSMT even ``against'' a significant bandwidth difference.

Our numerical results in the doped case show that the range of crystal field splitting $\Delta$ over which the orbital-selective Mott phases extend increases significantly with interaction $U$ (and Hund rule couplings $V=U/2$ and $J_z=U/4$). This can also be expected at half filling, in line with our analytical finding that the effective Heisenberg Hamiltonian relevant in this case and at strong coupling does not explicitly depend on $\Delta$, which is, consequently, also true for magnetic ordering temperatures. It would clearly be interesting to explore magnetic order also numerically away from half filling, possibly with orbital-dependent frustration.\cite{Lee2011a} However, quantitative accuracy would then require treatments beyond DMFT,\cite{Lee2011} at high numerical cost and probably with severe sign problems. 
In contrast, DMFT has been shown to be reliable for Mott physics as explored in this study; thus, our results are expected to be accurate for three-dimensional systems and experimentally relevant, e.g., at the lowest temperatures accessible with ultracold fermions on optical lattices.

\begin{acknowledgments}
  EJ was supported in part by the Graduate School of Excellence ``Materials Science in Mainz''. Support by the DFG within the TR 49 and by the John von Neumann Institute for Computing is gratefully acknowledged.
\end{acknowledgments}


\appendix

\section{Strong-coupling spectral functions at half filling}\label{sec:dos}

In this appendix we summarize the calculation of the spectral functions in the ground state ($T=0$) of the two-band model at strong coupling and half filling. The derivation proceeds along the lines of Kalinowski and Gebhard,\cite{Kalinowski2002} in which the {\it single\/}-band Hubbard model was studied in strong-coupling perturbation theory. Here we focus on the generic case $J_{\perp}< J_{z}$, which is somewhat simpler and, moreover, is the relevant model for the numerical calculations in this paper. Calculations for the special case $J_{\perp}= J_{z}=J$ proceed similarly, in principle, but they are technically more involved in detail.

We start with the definition of the local Green function $G_{m\sigma}(t)$, which contains (and is in fact equivalent to) the local spectral function:
\[
G_{m\sigma}(t)\equiv -\frac{i}{{\cal N}}\sum_{{\bf l}}\langle {\cal T}[ c_{{\bf l}m\sigma}^{\phantom{\dagger}}(t) c_{{\bf l}m\sigma}^\dagger (0)]\rangle_{\rm GS}\; .
\]
Here we introduced Heisenberg operators $c_{{\bf l}m\sigma}^{\phantom{\dagger}}(t)=e^{iKt} c_{{\bf l}m\sigma}^{\phantom{\dagger}}e^{-iKt}$
on site ${\bf l}$ for orbital $m$ and spin $\sigma$, where $K=H-\mu N$ is the grand canonical Hamiltonian. Moreover, ${\cal N}$ is the number of lattice sites, ${\cal T}$ the time ordering operator and $\langle \cdots \rangle_{\rm GS}$ an average over all possible degenerate $U=\infty$ ground states. The ground-state energy at half filling is denoted by $E_{0}$. The simplest way for imposing the restriction to half filling in the non-particle-hole symmetric {\it two-band\/} model $H_{\Delta}$ with $\Delta\neq 0$ is to construct a {\it four-band\/} model by duplicating the two-band Hamiltonian:
\[
 H_{\Delta}^{(1-4)}\equiv  H_{\Delta}^{(1,2)}+ H_{-\Delta}^{(3,4)}\; ,
\]
where the $(m=1,2)$- and $(m=3,4)$-orbitals have crystal field parameters $\Delta$ and $-\Delta$, respectively. The resulting four-band Hamiltonian is particle-hole symmetric under the transformation $ c_{{\bf i}m\sigma}^{\phantom{\dagger}}\rightarrow (-1)^{i} c_{{\bf i},m\pm2,\sigma}^{\dagger}$, so that the chemical potential of the four-band model is exactly given by $\mu =\frac{1}{2}U+U'-\frac{1}{2}J_{z}$. Results for the original two-band model are then obtained by simply restricting consideration to the $(m=1,2)$-orbitals in the end.

A Fourier transform from the time to a frequency variable yields two contributions, one from negative and one from positive frequencies, which are associated with the lower and upper Hubbard bands, respectively:
\[
G_{m\sigma}(\omega )=G_{m\sigma}^{\rm LHB}(\omega ) + G_{m\sigma}^{\rm UHB}(\omega ) \; .
\]
The contributions from the Hubbard bands are
\begin{eqnarray*}\nonumber
G_{m\sigma}^{\rm LHB}(\omega ) &=& \frac{1}{{\cal N}}\sum_{{\bf l}}\langle  c_{{\bf l}m\sigma}^\dagger \frac{1}{\omega +K-E_{0}-i0^{+}} c_{{\bf l}m\sigma}^{\phantom{\dagger}} \rangle^{\phantom{\dagger}}_{\rm GS}\\
G_{m\sigma}^{\rm UHB}(\omega ) &=& -\frac{1}{{\cal N}}\sum_{{\bf l}}\langle  c_{{\bf l}m\sigma}^{\phantom{\dagger}} \frac{1}{\omega +K-E_{0}-i0^{+}} c_{{\bf l}m\sigma}^\dagger \rangle^{\phantom{\dagger}}_{\rm GS}\nonumber
\end{eqnarray*}
and are connected to the spectral function according to
\[
\nu_{m\sigma}(\omega )=\nu_{m\sigma}^{\rm LHB}(\omega )+\nu_{m\sigma}^{\rm UHB}(\omega )
\]
with $\nu_{m\sigma}^{\rm LHB}(\omega )=\frac{1}{\pi}{\rm Im}[G_{m\sigma}^{\rm LHB}(\omega )]$ for the lower and $\nu_{m\sigma}^{\rm UHB}(\omega )=-\frac{1}{\pi}{\rm Im}[G_{m\sigma}^{\rm UHB}(\omega )]$ for the upper Hubbard band. It is sufficient to calculate the four contributions $\nu_{m\sigma}^{\rm LHB}(\omega )$ to the lower Hubbard band, since the results for the upper band then follow automatically from particle-hole symmetry, i.e., $\nu_{m\sigma}^{\rm UHB}(\omega )=\nu_{m\pm 2,\sigma}^{\rm LHB}(-\omega )$

In order to calculate $\nu_{m\sigma}^{\rm LHB}(\omega )$, we perform a canonical transformation to new particles: $c_{{\bf l}m\sigma}^\dagger = e^{S(\bar{c})}\bar{c}_{{\bf l}m\sigma}^\dagger e^{-S(\bar{c})}$, which (by definition) leaves the total number of double occupancies $H_{0}(\bar{c})$ of these new particles invariant. The result is:
\[
G_{m\sigma}^{\rm LHB}(\omega ) = \frac{1}{{\cal N}}\sum_{{\bf l}}\langle  \bar{c}_{{\bf l}m\sigma}^\dagger \frac{1}{\omega +\bar{K}-E_{0}-i0^{+}} \bar{c}_{{\bf l}m\sigma}^{\phantom{\dagger}} \rangle_{\rm \overline{GS}}
\]
where ${\rm \overline{GS}}$ denotes the ground state in terms of the new particles, and, moreover, 
\begin{equation}\label{defHtprimebarc}
\bar{K}=e^{-S(\bar{c})}K e^{S(\bar{c})}\equiv H_{\rm t}'(\bar{c})+H_{0}(\bar{c})-\mu\bar{N}
\end{equation}
and $\bar{N} = e^{-S(\bar{c})}N e^{S(\bar{c})}$. Note that the right-hand side of Eq.\ (\ref{defHtprimebarc}) {\it defines\/} the effective hopping $H_{\rm t}'(\bar{c})$ of the new particles. We now consider a single hole, i.e., the removal of a single particle with spin $\sigma$ from band $m$, in an otherwise half-filled assembly of states $d^{\dagger}_{{\bf i}\uparrow}|0\rangle$ and $d^{\dagger}_{{\bf i}\downarrow}|0\rangle$ with $d_{{\bf i}\sigma}=\bar{c}_{{\bf i}2\sigma}\bar{c}_{{\bf i}1\sigma}$, which (as argued in \refs{analytics}) are lowest in energy if $0\leq J_{\perp}<J_{z}$. In the following we also need single occupancies $s^{\dagger}_{{\bf i}m\sigma}|0\rangle$ with $s_{{\bf i}m\sigma}=\bar{c}_{{\bf i}m\sigma}$. The effective hopping $H_{\rm t}'(\bar{c})$ for a single hole then follows from standard Harris-Lange degenerate perturbation theory as
\[
H_{\rm t}'(\bar{c})= -t_{m}\sum_{({\bf ij})} s_{{\bf i}\bar{m}\sigma}^\dagger d_{{\bf j}\sigma}^\dagger d_{{\bf i}\sigma}^{\phantom{\dagger}} s_{{\bf j}\bar{m}\sigma}^{\phantom{\dagger}}
\]
 with $m=1,2,3,4$ corresponding to $\bar{m}=2,1,4,3$, respectively. As a result,
 \[
 \bar{K}={\textstyle{\frac{1}{2}}}(U+J_{z})-t_{m}\sum_{({\bf ij})} s_{{\bf i}\bar{m}\sigma}^\dagger d_{{\bf j}\sigma}^\dagger d_{{\bf i}\sigma}^{\phantom{\dagger}} s_{{\bf j}\bar{m}\sigma}^{\phantom{\dagger}}+{\cal O}\left(\frac{t^{2}}{U}\right)\; ,\vspace*{1ex}
 \]
 where the latter term is negligible in the strong-coupling limit $U/t\rightarrow \infty$. With the definition 
 \[
 z\equiv\omega +{\textstyle{\frac{1}{2}}}(U+J_{z})-{\textstyle{\frac{1}{2}}}\Delta (\delta_{m1}+\delta_{m4}-\delta_{m2}-\delta_{m3})-i0^{+}
 \]
 one then finds that, to this order in the $t/U$-expansion,
 \[
G_{m\sigma}^{\rm LHB}(\omega ) = \frac{1}{{\cal N}}\sum_{{\bf l}}\langle  d_{{\bf l}\sigma}^\dagger s_{{\bf l}\bar{m}\sigma}^{\phantom{\dagger}} \frac{1}{z+H_{\rm t}'(\bar{c})} s_{{\bf l}\bar{m}\sigma}^\dagger d_{{\bf l}\sigma}^{\phantom{\dagger}} \rangle_{\rm \overline{GS}}
\]
simply describes the hopping of a single hole through a random environment of $d^{\dagger}_{{\bf i}\uparrow}|0\rangle$ and $d^{\dagger}_{{\bf i}\downarrow}|0\rangle$ sites. The dependence of the right hand side on $z$ already shows that the lower Hubbard bands are centered around $\omega =-\frac{1}{2}(U+J_{z}\pm\Delta )$, with the sign of $\pm\Delta$ depending on the band index $m$.

In order to calculate $G_{m\sigma}^{\rm LHB}(\omega )$, we consider the series expansion
\begin{eqnarray*}\nonumber
G_{m\sigma}^{\rm LHB}(\omega ) &=& \frac{1}{z{\cal N}}\sum_{{\bf l}}\langle  d_{{\bf l}\sigma}^\dagger s_{{\bf l}\bar{m}\sigma}^{\phantom{\dagger}} \sum_{n=0}^{\infty}\left[ -\frac{H_{\rm t}'(\bar{c})}{z}\right]^{n} s_{{\bf l}\bar{m}\sigma}^\dagger d_{{\bf l}\sigma}^{\phantom{\dagger}} \rangle_{\rm \overline{GS}} \\ \nonumber
&=& \frac{1}{2z}\sum_{k=0}^{\infty}\left[ S(z)\right]^{k}=\frac{1}{2z[1-S(z)]}\; ,
\end{eqnarray*}
where $S(z)$ describes the contributions to the series of hole motions such that the hole does not return to site ${\bf l}$ between start and finish. The factor $\frac{1}{2}$ in the second step occurs, because site ${\bf l}$ has $d^{\dagger}_{{\bf l}\sigma}|0\rangle$-character only with probability $\frac{1}{2}$. Furthermore, in each hop only {\it half\/} the sites are accessible, because the hole must hop to a $d^{\dagger}_{{\bf l}\sigma}|0\rangle$-site. Hence, on a Bethe lattice (with coordination $Z$ and $t_{m}^{*}\equiv t_m\sqrt{Z}$),
\[
S(z)={\textstyle{\frac{1}{2}}}Z\left(\frac{t_{m}^{*}}{z\sqrt{Z}}\right)^{2}\sum_{k'=0}^{\infty}\left[ S(z)\right]^{k'}=\frac{(t_{m}^{*})^{2}}{2z^{2}[1-S(z)]}\; ,
\]
yielding
\[
G_{m\sigma}^{\rm LHB}(\omega )=\frac{1}{2(t_{m}^{*})^{2}}\left[ z+\sqrt{z^{2}-2(t_{m}^{*})^{2}} \right]\; ,
\]
where the sign of the square root is chosen such that $G_{m\sigma}^{\rm LHB}(\omega )\sim \frac{1}{z}$ for $\omega\rightarrow -\infty$. For the $(m=1)$- and $(m=2)$-orbitals, this immediately yields the results for the spectral functions quoted in \refs{analytics} for $J_{\perp}< J_{z}$.

As mentioned above, calculations for the special case $J_{\perp}= J_{z}=J$ are analogous but considerably more involved, so that we prefer to simply quote the result (see \refs{analytics}).


\bibliography{ref_noURLs.bib,nils.bib,publications_Bluemer.bib}

\end{document}